\documentstyle[aps,twocolumn,epsfig,floats,prl,times]{revtex}

\begin{document}
\draft


\title{Strong-field terahertz-optical mixing in excitons}
\author {M. Y. Su, S. Carter, and M. S. Sherwin}
\address {
Physics Department and Center for Terahertz Science and Technology,
University of California, Santa Barbara, California 93106}
\author {A. Huntington and L. A. Coldren}
\address {
Materials Department, University of California,
Santa Barbara, California 93106}

\date{\today}
\maketitle
\begin{abstract}

Driving a double-quantum-well excitonic intersubband resonance
with a terahertz (THz) electric field of frequency $\omega_{THz}$
generated terahertz optical sidebands
$\omega=\omega_{THz}+\omega_{NIR}$ on a weak NIR probe.  At high
THz intensities, the intersubband dipole energy which coupled two
excitons was comparable to the THz photon energy.  In this
strong-field regime the sideband intensity displayed a
non-monotonic dependence on the THz field strength.  The
oscillating refractive index which gives rise to the sidebands may
be understood by the formation of \em Floquet states\em, which
oscillate with the same periodicity as the driving THz field.

\end{abstract}

\pacs{78.70.-g,42.65.-k,73.21.Fg}

\narrowtext

Electromagnetically-induced coherent quantum effects in
semiconductor systems include Rabi oscillations in
donors\cite{Cole01} and quantum dots\cite{Stievater01}, and
electromagnetically-induced transparency in quantum
wells\cite{Serapiglia00}.  In these effects, to induce coherence
the driving field must overcome dephasing processes. Thus, the
Rabi frequency $\mu E$ must be greater than the dephasing rates
$\gamma$: $\mu E > \hbar\gamma$, where $\mu$ is the dipole moment
and $E$ is the electric field strength.

If the field strength is increased still further, a \em
strong-field \em regime is encountered as the Rabi energy becomes
comparable to the photon energy: $\mu E \geq \hbar\omega$.
Previous studies into this regime have looked at atoms in
microwave cavities\cite{Leeuwen1985,Haffmans1994} and strong laser
fields\cite{Gavrila92}. Since atomic excited states are near the
continuum, strong-field effects manifest themselves as
nonmonotonic multi-photon ionization rates as a function of field
strength.

However, the atoms are ionized, meaning the bound states do not
exist after the microwave or laser field is turned on.
Multi-photon ionization is an inescapable result of driving
excited atomic levels which get closer and closer together.

There is a rich body of theoretical predictions on the effect of
strong fields on the \em bound \em states of a quantum
system\cite{Johnsen1999,Fromherz1997}. In quantum wells, upper
states get further and further apart.  Thus the strong-field
condition can be satisfied by resonantly driving the lowest
subbands of a quantum well at THz frequencies.  Yet the quantum
well is far deeper than the Rabi energy, so the bound states still
survive.

We describe experiments in which a near-infrared (NIR) probe laser
beam is mixed with a terahertz (THz) pump beam in a gated,
asymmetric double quantum well (DQW). The THz field couples to an
excitonic \em intersubband \em excitation while the NIR field
couples to an excitonic \em interband \em excitation. Applying a
DC voltage to the gates brings the intersubband transition into
resonance with the THz field, and the interband transition into
resonance with the NIR field.  When these resonance conditions are
met the NIR probe is modulated resulting in the emission of
optical sidebands which appear at frequencies $\omega_{sideband} =
\omega_{NIR} + n\omega_{THz}$ where $\omega_{NIR}$
($\omega_{THz}$) is the frequency of the NIR (THz) beam and $n=\pm
1,2,\ldots$. Strong-field effects manifest themselves as
non-monotonic sideband intensities as a function of THz field
strength.

The sample consisted of an active region, gates, and a distributed
Bragg reflector (DBR).  The active region consisted 5 periods of
DQW, each consisting of a 120 {\AA} GaAs QW and a 100 {\AA} GaAs
QW separated by a 25 {\AA} Al$_{0.2}$Ga$_{0.8}$As tunnel barrier.
Each period was separated by a 300 {\AA} Al$_{0.3}$Ga$_{0.7}$As
barrier.  The active region was sandwiched between two gates, each
consisting of a Si delta-doped 70 {\AA} QW with carrier density
$\approx 1\times 10^{12}$ cm$^{-2}$. The gates were separated from
the active region by 3000 {\AA} Al$_{0.3}$Ga$_{0.7}$As barriers.
Since the gate QWs are much narrower than the active DQWs, the
gate QWs are transparent to both the THz and NIR beams. The gated
DQWs were grown on top of a DBR which consisted of 15 periods of
689 {\AA} AlAs and 606 {\AA} Al$_{0.3}$Ga$_{0.7}$As. It had a
low-temperature passband nearly centered on the low-temperature
bandgap of the DQW, making it about 95\% reflective for the NIR
probe beam and sidebands.

\begin{figure}
\centerline{\epsfig{file=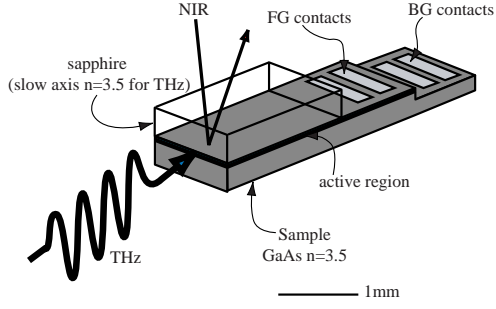,width=0.75\linewidth}}
\caption {Processed waveguide device and optical coupling scheme.}
 \label{waveguide}
\end{figure}

We etched a mesa and annealed NiGeAu ohmic contacts to the gate
QWs.  The sample was then cleaved into a 1 mm wide strip, 8 mm
long.  We then cleaved a 400 $\mu$m-thick wafer of crystal
sapphire substrate material into a $1\times5$ mm strip, oriented
with the optically slow axis along the long dimension. The
sapphire was mechanically pressed against the surface of the
sample with a beryllium copper clip as shown in Fig.
\ref{waveguide}.  At cryogenic temperatures sapphire is
index-matched to GaAs and transparent at THz wavelengths.  At the
same time, sapphire is transparent to the NIR probe.  Pressing the
sapphire against the sample forms a rectangular dielectric
waveguide with half of the waveguide defined by the sample
substrate and the other half defined by the sapphire.  The
epilayer containing the active region lies in the middle of the
heavily over-moded waveguide.  This results in much higher fields
at the active region than would be possible without the sapphire,
where the active region would be at the edge of a dielectric
waveguide.

The sample was cooled to 21 K in a closed-cycle He cryostat. The 1
kW THz beam from a free-electron laser is polarized in the growth
direction, propagated in the QW plane and focused by a 90$^o$
off-axis parabolic mirror (F/1) into the edge of the waveguide.
The maximum THz electric field strength at the focus is estimated
to be between $\approx 5-20$ kV/cm, where our uncertainty comes
from uncertainty of the exact size of the mode in the waveguide.
The THz power can be continuously varied by a pair of wire-grid
polarizers.

NIR light from a continuous-wave Ti:sapphire laser is chopped by
an acousto-optic modulator into 25 $\mu$s pulses which overlap the
5 $\mu$s FEL pulses at maximum repetition rate of 1.5 Hz. The
vertically-polarized NIR beam propagates normal to the THz beam,
and was focused (F/10) at a power density of $\approx$ 5 W/cm$^2$
to the same small interaction volume in the sample.  The reflected
beam, sidebands, and photoluminescence (PL) are analyzed by a
second polarizer, dispersed by a 0.85 m double-monochromator, and
detected by a photomultiplier tube.

\begin{figure}
\centerline{\epsfig{file=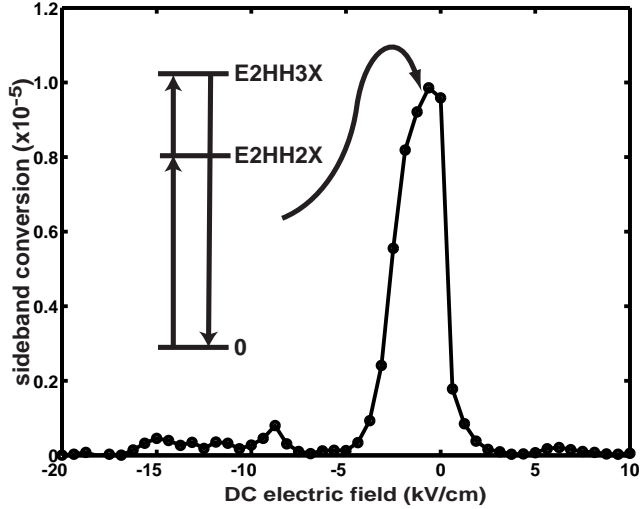,width=1.0\linewidth}}
\caption {The dc electric field dependence of sideband at
$\hbar\omega_{NIR}=1547$ meV and $\omega_{THz}=1.5$ THz (6.2 meV).
The inset illustrates the resonant condition in which the NIR
field is resonant with the E2HH2 exciton and the THz field
resonantly couples the E2HH2 and E2HH3 excitons.} \label{sbvs}
\end{figure}

Previous work \cite{su1_02,su2_02} showed that sideband generation
is enhanced when the NIR field resonantly couples the vacuum state
$|0\rangle$ with an exciton $|1\rangle$, and the THz field
resonantly couples with the intersubband transition between
excitons $|1\rangle$ and $|2\rangle$.  The exciton states can be
labelled as E$_\mu$HH$_\nu$X, indicating an exciton consisting of
an electron in conduction subband $\mu$ and a heavy hole in
valence subband $\nu$.  Changing the DC electric field tunes the
excitonic intersubband transition into resonance with the THz
field, as shown in Fig. \ref{sbvs}.  The sideband resonance we
focused on is E2HH2X-E2HH3X, a peak assignment made by comparing
low-field sideband spectroscopy with a nonlinear susceptibility
calculation for excitons\cite{su2_02}.

\begin{figure}
\centerline{\epsfig{file=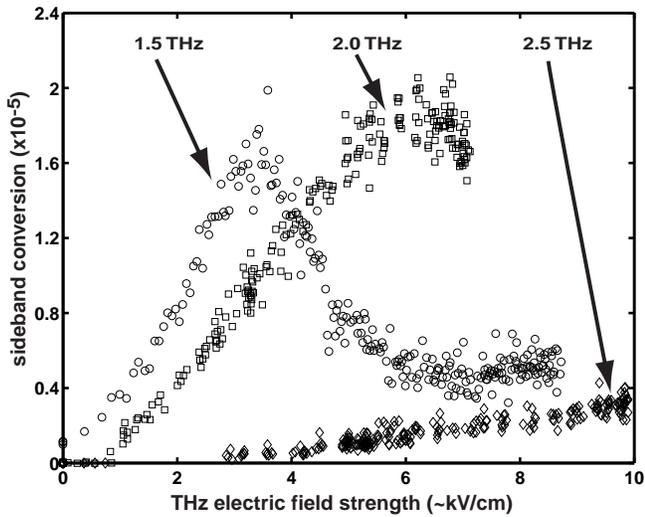,width=1.0\linewidth}}
\caption {Resonant THz field strength dependence of sideband
generation at 1.5 THz, 2.0 THz, and 2.5 THz (6.2 meV, 8.2 meV,
10.4 meV).  The absolute THz electric field scale is accurate only
to within a factor of 2.} \label{pd506684}
\end{figure}

Fig. \ref{pd506684} shows the dependence on THz field strength of
sideband generation at various THz frequencies.  Each curve was
taken at a gate bias near the E2HH2X-E2HH3X
resonance\cite{su2_02}. The most striking feature is the
non-monotonic behavior in the strong-field regime when the Rabi
energy is comparable to the photon energy.  Clearly at lower
frequencies the power dependence rolls over at a lower field
strength.

\begin{figure}
\centerline{\epsfig{file=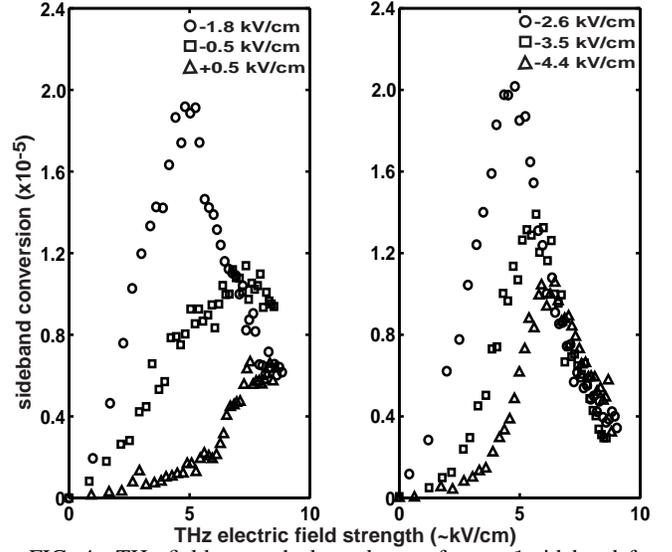,width=1.0\linewidth}}
\caption {THz field strength dependence of $n=1$ sideband for
$\omega_{NIR}$=1548 meV, $\omega_{THz}=1.5$ THz (6.2 meV) at
various DC electric field detunings from the resonance condition
illustrated in Fig. \ref{sbvs}.  The absolute THz electric field
scale is accurate only to within a factor of 2.}
\label{powerdep}
\end{figure}

By sitting at the peak NIR frequency for E2HH2X-E2HH3X and varying
the DC gate voltage we took THz power dependence scans at various
THz detunings $E_2-E_1-\hbar\omega_{THz}$ where $E_2$ ($E_1$) is
the energy of the upper (lower) exciton state.  The results are
shown in Fig. \ref{powerdep}. As the detuning is varied, so does
the shape of the THz field dependence.

The power dependence cannot be explained by a nonlinear
susceptibility, which is inherently a low-field theory because it
relies on a Taylor expansion about the field strength \cite{Boyd}.
Saturation effects\cite{Heyman1994} in a nonlinear susceptibility
can only come about when contributions from virtual transitions
initiated from excited states $|1\rangle$ or $|2\rangle$ are
comparable to those initiated from the vacuum $|0\rangle$. However
the populations of excited states $|1\rangle$ or $|2\rangle$ are
never significant compared with $|0\rangle$ in our undoped sample
and weak NIR beam. Therefore a nonlinear susceptibility can only
predict a linear (quadratic) dependence of sideband intensity on
THz power (field strength).

The following simple, phenomenological, three-state model captures
qualitative features of the dependence of sidebands on THz power.
The Hamiltonian $H_0$ has eigenstates $\phi_0(x,z)$,
$\phi_1(x,z)$, and $\phi_2(x,z)$ with eigenenergies $E_0=0$,
$E_1$, and $E_2$, respectively.  Near the experimental resonance,
the effect of the DC voltage is to tune $\phi_2$ while the other
$\phi_1$ remains relatively unchanged\cite{su2_02}. The two
exciton states $\phi_1(x,z)$ and $\phi_2(x,z)$ are coupled by the
$z$-dipole operator for the DC and THz fields.  The two upper
states $\phi_1(x,z)$ and $\phi_2(x,z)$ are coupled to the ground
state by the $x$-dipole operator for the NIR field.

Therefore, the x and z matrices are
\begin{eqnarray}
\nonumber z=\left({
\begin{array}{ccc}
0 & 0 & 0 \\
0 & 0 & z_{12} \\
0 & z_{21} & z_{22}
\end{array}}\right)
x=\left({
\begin{array}{ccc}
0 & x_{01} & x_{02} \\
x_{10} & 0 & 0 \\
x_{20} & 0 & 0
\end{array}}\right)
\end{eqnarray}
All the nonzero terms
$x_{\alpha\beta}=\langle\phi_\alpha|x|\phi_\beta\rangle$ and
$z_{\alpha\beta}=\langle\phi_\alpha|z|\phi_\beta\rangle$ are set
equal to unity. The Hamiltonian is
\begin{displaymath}
H=H_0+zE_\omega\cos{\omega t}+x\lambda E_\Omega\cos{\Omega t}
\end{displaymath}
where $E_\omega$ represents the strong THz electric field strength
and $E_\Omega$ represents the weak NIR electric field.  We solved
the Hamiltonian nonperturbatively within a Floquet formalism for
$\lambda=0$, while the weak probe was added later using
time-dependent perturbation theory.

The solutions to the Schrodinger equation for the time-periodic
Hamiltonian have the form\cite{Shirley1965}
\begin{eqnarray}
\nonumber \varphi_i(z,t)=e^{-i\epsilon_i t/\hbar}u_i(z,t),
u(z,t)=u(z,t+\frac{2\pi}{\omega})
\end{eqnarray}
where $u_i(z,t)$ has the same periodicity as the driving frequency
and and $\epsilon_i$ is called the \em quasienergy\em.  The states
$\varphi_i(z,t)$ are called \em Floquet states \em and are
mathematically analogous to Bloch states for a spatially-periodic
Hamiltonian.  Meanwhile the quasienergies are analogous to the
crystal momenta of Bloch theory.

The spatial-dependence of the wavefunction $\varphi_i(z,t)$ can be
expanded in terms of the original eigenstates $\phi_\alpha(z)$,
and the time-dependence can be Fourier-expanded
\begin{equation}
\varphi_i(z,t)= e^{-i\epsilon_i
t/\hbar}\sum_{\alpha,n}c^i_{\alpha,n}e^{-in\omega t}\phi_\alpha(z)
\label{eq_floquetexpansion}
\end{equation}
The coefficients $c^i_{\alpha,n}$ are what will determine the part
of the index of refraction which will oscillate at particular
multiples of $\omega$.  Solving for these coefficients is the key
to understanding the power dependence of the sideband.  These
Floquet coefficients are closely related to the photon-assisted
tunnelling sidebands which appear in the irradiated
current-voltage curves in superconducting weak-link
junctions\cite{TienGordon} and coupled quantum
wells\cite{Drexler95}.


The time-dependent Schrodinger equation can be cast in terms of
the Floquet operator $F=H- i\hbar\frac{\partial}{\partial t}$,
which can be written as a matrix of the form\cite{Fromherz1997}
\begin{equation}
F=\left({
\begin{array}{ccccc}
  \ddots&[\mu]&0&0&0\\
  {[\mu]}&[H_0]-\hbar\omega&[\mu]&0&0 \\
  0&[\mu]&[H_0]&[\mu]&0 \\
  0&0&[\mu]&[H_0] + \hbar\omega&[\mu] \\
  0&0&0&[\mu]&\ddots
\end{array}}\right)
\label{eq_floquetmatrix}
\end{equation}
where $[H_0]$ and $[\mu]$ are matrix representations of the
operators $H_0$ and $\mu$ in the $|\phi_\alpha\rangle$ basis. In
practice the Floquet matrix must be truncated up to $\pm N$
photons.  $N$ can be made arbitrarily large for an arbitrarily
precise result at high field strengths. The solution to the
Schrodinger equation for a time-periodic Hamiltonian reduces to
finding the eigenvalues and eigenvectors of the Floquet matrix
(\ref{eq_floquetmatrix}).  Other than the $\pm N$ truncation the
solution involves no perturbation or rotating-wave approximations.
Here we used $N=16$ photons.

The Floquet solutions are labeled $\varphi_i(x,z,t)$ with
$i=0,1,2$, and expanded as in (\ref{eq_floquetexpansion}).  Their
expansions are explicitly expressed for clarity:
\begin{eqnarray*}
\varphi_0(x,z,t)&=&\phi_0(x,z)\\*
\varphi_1(x,z,t)&=&e^{i\epsilon_1 t/\hbar}\sum_{n}e^{-in\omega
t}[c^1_{1,n}\phi_1(x,z)+c^1_{2,n}\phi_2(x,z)]\\*
\varphi_2(x,z,t)&=&e^{i\epsilon_2 t/\hbar}\sum_{n}e^{-in\omega
t}[c^2_{1,n}\phi_1(x,z)+c^2_{2,n}\phi_2(x,z)]
\end{eqnarray*}
The ground state $\phi_0(x,z)$ is not coupled by the the strong
field to the 2-d subspace spanned by $|\phi_1\rangle$ and
$|\phi_2\rangle$. Therefore the coefficients for the Floquet state
$\varphi_0(x,z,t)$ vanish except $c^0_{0,0}=1$.  In other words,
the ground vacuum state remains untouched by the intense THz
field.

The dipole response of the Floquet states to the weak NIR probe is
calculated perturbatively. The state of the system with both
driving fields $\psi(x,z,t)$ can be expanded in terms of the
Floquet states $\varphi_i(x,z,t)$ to first order.  Discarding
second order terms and anti-resonant contributions, the dipole
expectation is given by
\begin{equation}
\langle \psi|x|\psi \rangle = E_\Omega e^{-i\Omega t}\sum_{i=1}^2
\frac{\langle\varphi_0|x|\varphi_i\rangle
\langle\varphi_i|x|\varphi_0\rangle} {\epsilon_i-\hbar\Omega}
\label{eq_floquetexp}
\end{equation}
The form of (\ref{eq_floquetexp}) is the same as the linear
susceptibility of an undriven system, except the states
$\varphi_i$ are oscillating Floquet states instead of stationary
states.  Explicitly expanding the Floquet states in the numerator
gives us an expression for the polarization
\begin{equation}
x(t)= E_\Omega e^{-i\Omega t}
\sum_{i=1}^2\mathop{\sum_{n,\alpha}}_{m,\beta}
 \frac
{c^i_{n,\alpha}c^i_{m,\beta}e^{-i(n-m)\omega t} x_{0\beta}
x_{\alpha 0} } {\epsilon_i-\hbar\Omega}
\label{eq_polarization}
\end{equation}

\begin{figure}
\centerline{\epsfig{file=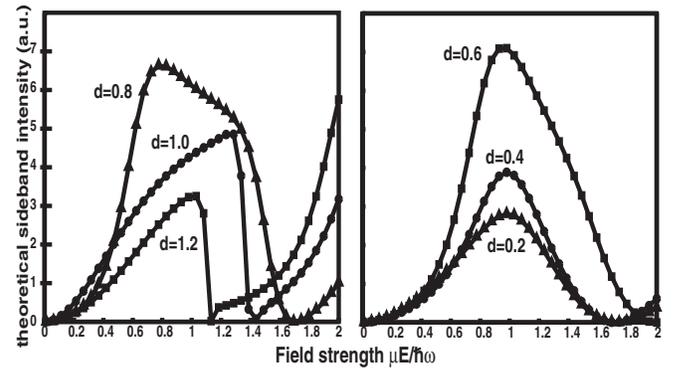,width=1.0\linewidth}}
\caption {Sidebands calculated by evaluating the square of
(\ref{eq_floquet_sb}) at various detuning parameters LEFT:
$d=\frac{E2-E1}{\hbar\omega}=0.8,1.0,1.2$, RIGHT:
$d=\frac{E2-E1}{\hbar\omega}=0.2,0.4,0.6$.}
 \label{sbtheory}
\end{figure}

The $+1\omega$ sideband is given by the Fourier component of
$x(t)$ oscillating at the frequency $\Omega+1\omega$.  Also, for
the resonant condition illustrated in Fig. \ref{sbvs},
$\hbar\omega\approx\epsilon_1$, so we keep only the $i=1$ term in
the sum (\ref{eq_polarization}).  This condition is satisfied when
$n-m=1$. Thus we obtain an expression for the sideband
polarization
\begin{equation}
x_{sideband}(t)=\frac{E_\Omega e^{-i(\Omega + \omega)
t}}{\epsilon_1-\hbar\Omega}\sum_{n,\alpha,\beta}c^1_{n+1,\alpha}c^1_{n,\beta}
x_{0\beta} x_{\alpha 0} \label{eq_floquet_sb}
\end{equation}
The intensity of the sideband is proportional to $x_{sideband}^2$.
This is plotted vs. field strength for various detunings in Fig.
\ref{sbtheory}.  The detuning parameter $d$ is the level spacing
normalized by the photon energy of the strong field,
$d=\frac{E2-E1}{\hbar\omega}$.

Our model captures the rollover of the resonant power dependence
up to field strengths of around $\frac{\mu E}{\hbar\omega}\approx
1.5$.  Given a calculated excitonic intersubband dipole of
$\frac{\mu}{e}=$ 12 nm\cite{su2_02}, the strong-field condition is
met at a THz field of 5 kV/cm.  However, experimentally the
sideband never completely disappears, a striking prediction of the
theory. It is unlikely that a 3-state simplification is entirely
valid because there is a spectrum of other exciton states as well
as an electron-hole continuum that may provide significant
off-resonant contributions.

To summarize, we drove an quantum-well excitonic intersubband
transition with an intense THz laser field in a regime where the
Rabi energy was comparable to the photon energy.  The nonlinear
mixing between the THz pump and a weak NIR probe displayed a
nonmonotonic dependence on THz power, which could not be explained
by a conventional nonlinear susceptibility.  Rather, the excitons
were dressed by the THz field, a process which we described in a
simple three-level model in which the interaction between excitons
and the THz field is solved non-perturbatively within the Floquet
formalism.

This research is funded by NSF-DMR 0070083.

\bibliography{/mark_su/bibtex/su_aps}
\bibliographystyle{unsrt}

\clearpage

\end{document}